\documentclass[aps,prl,preprint,superscriptaddress,nobibnotes]{revtex4-2}
\usepackage{graphicx,color}
\usepackage{amsmath}
\usepackage{bm}
\usepackage{dcolumn}
\usepackage{ifthen}
\usepackage{multirow}
\usepackage{threeparttable}
\usepackage{siunitx}
\usepackage{hyperref}
\hypersetup{backref=true, pdfnewwindow=true, linkcolor=blue, colorlinks=true, anchorcolor=blue, citecolor=blue, filecolor=blue,  menucolor=blue, urlcolor=blue}

\begin{document}
\title{Hydrogen toggling between Yoshimori spin spirals and elliptical Dzyaloshinskii-Moriya skyrmions in Fe on Ir(110)}

\author{Timo Knispel}
\affiliation{II. Physikalisches Institut, Universit\"{a}t zu K\"{o}ln, Z\"{u}lpicher Str. 77, 50937 Cologne, Germany \looseness=-1}
\author{Vasily Tseplyaev}
\affiliation{Peter Gr\"{u}nberg Institut, Forschungszentrum J\"{u}lich and JARA, D-52425 J\"{u}lich, Germany \looseness=-1}
\affiliation{Physics Department, RWTH-Aachen University, D-52062 Aachen, Germany \looseness=-1}
\author{Gustav Bihlmayer}
\affiliation{Peter Gr\"{u}nberg Institut, Forschungszentrum J\"{u}lich and JARA, D-52425 J\"{u}lich, Germany \looseness=-1}
\author{Stefan Bl\"ugel}
\affiliation{Peter Gr\"{u}nberg Institut, Forschungszentrum J\"{u}lich and JARA, D-52425 J\"{u}lich, Germany \looseness=-1}
\affiliation{Physics Department, RWTH-Aachen University, D-52062 Aachen, Germany \looseness=-1}
\author{Thomas Michely}
\affiliation{II. Physikalisches Institut, Universit\"{a}t zu K\"{o}ln, Z\"{u}lpicher Str. 77, 50937 Cologne, Germany \looseness=-1}
\author{Jeison Fischer}
\email{jfischer@ph2.uni-koeln.de}
\affiliation{II. Physikalisches Institut, Universit\"{a}t zu K\"{o}ln, Z\"{u}lpicher Str. 77, 50937 Cologne, Germany \looseness=-1}

%\date{\today}

\begin{abstract}
Skyrmions are particle-like spin textures that arise from spin spiral states in the presence of  an external magnetic field. These spirals can originate from either frustrated Heisenberg exchange interactions or the interplay between exchange interactions and the relativistic Dzyaloshinskii-Moriya interaction, leading to  atomic- and mesoscale textures, respectively. However, the conversion of exchange-stabilized spin spirals into skyrmions typically requires magnetic fields that exceed practical laboratory limits. Here, we demonstrate a strategy leveraging hydrogen adsorption to expand the range of magnetic films capable of hosting stable or metastable skyrmions. In a structurally open and anisotropic system of two pseudomorphic Fe layers on Ir(110), spin-polarized scanning tunneling microscopy combined with ab initio calculations reveals that a right-handed, exchange-stabilized Néel-type spin spiral propagating along the [$\overline{1}10$] direction with a $1.3$~nm period transitions upon hydrogen adsorption to a Dzyaloshinskii-Moriya type spiral with a sevenfold longer period of $8.5$~nm. This transition enables elliptical skyrmions to form at moderate magnetic fields. Hydrogenation thus provides a non-volatile mechanism to toggle between distinct magnetic states, offering a versatile platform for controlling spin textures. 

\end{abstract}

\maketitle
\newpage
Over the past 15 years, the understanding, development, and manipulation of spatially localized magnetization textures~\cite{Goebel2021} with length scales, thermal and dynamic stabilities, lifetimes and transport properties compatible with potential technological applications has become an extremely active field of research~\cite{Fert2013,Krause2016,Back2020,Yang2021,Song2020,Everschor-Sitte2024,Chen2024,Larraga:2024}. Among the magnetization textures, skyrmions~\cite{Bogdanov:1989.1,Roessler2006} (as well as antiskyrmions~\cite{Hoffmann2017}), particles with a two-dimensional topologically non-trivial magnetization texture, play a pivotal role. It is well-established from dimensional analysis that skyrmions cannot be stabilized solely combining ferromagnetic (FM) Heisenberg interactions and magnetic anisotropy. In particular, Bogdanov \textit{et al.}~\cite{Bogdanov:1989.1} demonstrated that the Dzyaloshinskii-Moriya interaction~\cite{Dzyaloshinskii1957, Moriya1960} (DMI), which which breaks the degeneracy between chiral states, serves as a crucial stabilizing interaction for these topological structures. For weak DMI, the single skyrmion appears as a metastable state in the presence of an external magnetic field. In the regime of strong DMI, the ferromagnetic ground state is replaced by a chiral (Dzyaloshinskii~\cite{Dzyaloshinskii1965}) spin-spiral state. As the external field increases, a skyrmion lattice emerges, followed by the stabilization of individual skyrmions~\cite{Bogdanov2020}.

Frustrated exchange, arising from competing antiferromagnetic (AFM) and FM Heisenberg interactions between different atomic pairs, offers an alternative, achiral mechanism for skyrmion stabilization~\cite{Leonov2015}. Achirality implies that skyrmions and antiskyrmions have degenerate energies. Given that the exchange interaction is significantly stronger than the relativistic DMI, the resulting (Yoshimori~\cite{Yoshimori1959}) spiral ground states exhibit much shorter pitches, which should lead to reduced skyrmion sizes -- an advantageous property for various applications. This has prompted extensive investigations in recent years, particularly in rare-earth intermetallics~\cite{Kurumaji2019, Paddison2022, khanh2020, hirschberger2019, Takagi2022, Moya2022}. However, the stronger frustrated exchange interaction presents a limitation: while laboratory magnetic fields can still transform the spiral state into a skyrmion lattice phase, they are insufficient to stabilize individual skyrmions. In transition-metal films, such as Fe/Cu(111)~\cite{Phark2014}, the frustrated exchange is so strong that even skyrmion lattice phases cannot be stabilized within the limits of laboratory magnetic fields.

A common strategy to modify the spin texture of ultrathin films involves adding a heavy metal overlayer to tune the DMI at the new interface~\cite{Dupe2016,Moreau-Luchaire2016,Velez2022}. Alternatively, electric field gating has emerged as a promising technique to dynamically control DMI and magnetic anisotropy in ultrathin magnetic materials by inducing charge distribution or ionic migration~\cite{Srivastava2018}. However, many of these effects rely on volatile mechanisms, \textit{e.g.}, where the modified state persists only while the electric field is applied. For practical memory and logic applications, achieving non-volatile control over magnetic interactions is essential to ensure energy-efficient and persistent manipulation of spin textures~\cite{HerreraDiez2019,Wang2020}. Surface gas functionalization, particularly with hydrogen, offers a promising route. Hydrogen, readily adsorbed on metal surfaces, can affect magnetic properties~\cite{Havela2022} like the magnetic moment, anisotropy~\cite{Sander2004,Santos2012}, and induce subtle DMI~\cite{Chen2021}, leading to skyrmion formation via magnetic anisotropy~\cite{Chen2022}. While hydrogen has been used to induce skyrmions in Fe films on Ir(111), yet an irreversible structural transition appears to accompany the hydrogen adsorption~\cite{Hsu2018}. To reversibly adjust frustration in a non-volatile manner, hydrogen must modify solely the exchange interaction via changes in the local electronic structure.

In this Letter, we make use of the openness of the fcc(110) surface and demonstrate for the case of double layer Fe on Ir(110) that hydrogen adsorption enables a transition from a short-pitch Yoshimori spiral to a long-pitch DMI spiral. The DMI spiral can be further transformed into single skyrmions within a background of field-aligned spins under an external magnetic field. Due to the $C_{2v}$ symmetry of the surface, the exchange and, to a lesser extent, the DMI are anisotropic, leading to elliptical skyrmions.

Our recent spin-polarized scanning tunneling microscopy (STM) study of ultrathin Fe films on Ir(110) have demonstrated that Fe grows pseudomorphically as double layer and exhibits a right-handed cycloidal atomic-scale spin spiral with a period of about 1.3~nm. The spin spiral is robust and an external magnetic field of up to 9~T is not able to unwind it~\cite{Knispel2025}. Density functional theory (DFT) calculations identified the spiral as of Yoshimori-type resulting from the competition of a FM Heisenberg exchange interaction between interplane nearest neighbor atom pairs and the AFM intraplane pair interaction of  more-distant neighbors leading in total to a frustrated Heisenberg exchange system whose ground state is a spin spiral with an energy gain of 12~meV/Fe atom with respect to the FM state.  The DMI selects  a Néel over a Bloch spiral and selects the unique handedness, but leaves the period largely unchanged.

Exchange interactions $J(\varepsilon)$ depend on the band energy $\varepsilon$ (see, \textit{e.g.}, Ref.~\onlinecite{Terakura1982}) and changing the band filling is a path to changing their strength and sign and subsequently the magnetic order. This is particularly effective for frustrated magnets, where the magnetic order depends on competing exchange interactions of opposite signs between atom pairs at varying distances. Alloying for example is a very well-known concept making use of this phenomenon~\cite{Ferriani2007}. An alternative approach on a much finer scale is the change of the band occupation by moving the Fermi energy and thereby varying the number of electrons in the system. A practical realization is the absorption of hydrogen (H) on the system. It leaves the Fe and Ir bands largely unchanged but, dependent on coverage, shifts the Fermi energy relative to the $d$-states by a fraction of an eV. Adsorption thus changes band occupation. Thereby the magnetic order can be wound up into a more interesting tunable multi-Q texture using magnetic fields accessible in laboratory environments.

%Figure 1
\begin{figure}[ht!]
\includegraphics[width=0.45\textwidth]{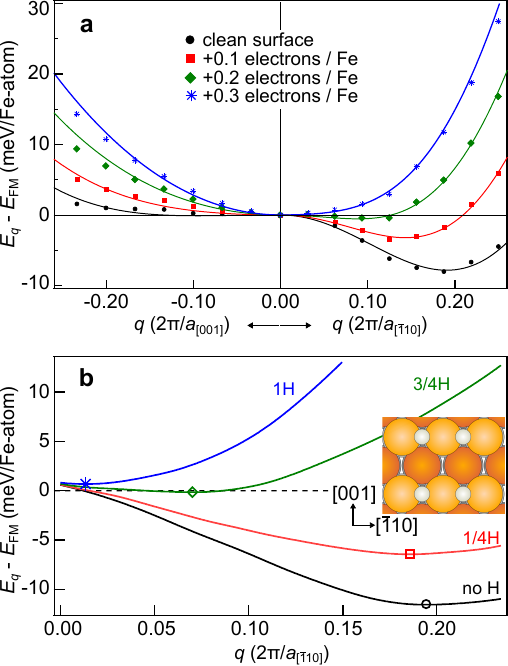}
\caption{DFT calculated energetics of spin spirals.
(a) Energy dispersion $E(q)$ of clean 2\,ML Fe/Ir(110) along two orthogonal directions and with virtually additional charge of 0.1~e, 0.2~e and 0.3~e. The calculations were performed scalar-relativistically with LDA using the structure optimized in GGA shown in Note~1 in the SI.
(b) Energy dispersions $E(q)$ with SOC relative to the FM state of homogeneous, flat cycloidal spin spirals as a function of the wave vector $q$ for 2~ML Fe/Ir(110). Hydrogenated films for different H concentration (given in the graph). The inset displays a ball model with the adsorbed H at full coverage as white spheres.
}
\label{dft}
\end{figure}

We test these hypotheses theoretically by conducting vector-spin DFT  calculations using the film version of the full-potential linearized augmented plane wave method~\cite{wimmer1981flapw} (FLAPW) as implemented in the FLEUR code~\cite{fleur2023}, to analyze the magnetic order and energetics of the magnetic structure through the calculations of homogeneous magnetic spin-spirals in (i) a  pristine Fe bilayer film on an 11-layer Ir substrate modeling the experimental sample by increasing the Fermi energy relative to the $d$-band through the addition of  charges of 0.1, 0.2 and 0.3~e/Fe atom by increasing the nuclear number of Fe accordingly and (ii)  by this film absorbed with different coverages of H. For the systems with H, the calculations were performed in a $p(2\times2)$ unit cell. Several H adsorption sites in the unit cell were tested. The energetically most stable configuration was identified with H located in a bridge position between the topmost Fe atoms. The atomic structure has been optimized for each H coverage (see Note~1 in the SI).  The calculations  of spin spirals have been performed incorporating configurations with one, three or four hydrogen atoms at distinct  bridge positions (see white spheres in the inset of Figure~\ref{dft}b), corresponding to 0.25\,ML, 0.75\,ML and 1.0\,ML of H on Fe/Ir(110), exploiting the generalized Bloch theorem~\cite{Kurz:04} for calculations without SOC. The contribution of SOC to the spirals has been treated in first order perturbation theory~\cite{Heide:09}. For the calculation of the magnetic anisotropy, SOC was included self-consistently~\cite{Li:90}. The local density approximation (LDA)~\cite{Vosko:80} was applied, resulting in an optimized Ir lattice constant of $a_0=0.382$~nm in good agreement to experiment (0.3839~nm). 

The results are shown in Figure~\ref{dft}. Figure~\ref{dft}a illustrates that the addition of extra charge to the Fe layer, which occupies previously unoccupied Fe and Ir states, results in a significant change of the exchange interactions demonstrated in terms of the energy dispersion of the spin-spiral as a function of the wave vector $q$. Specifically, the balance of competing FM and AFM exchange interactions along the $[\bar{1}10]$ directions shifts towards ferromagnetism. Figure~\ref{dft}b reproduces this behavior, but as function of the H concentration. Increasing the H concentration increases the charge transfer from H to the Fe/Ir system and the energy dispersion as function of the wave vector $\mathbf {q}$ shows a significant change in the energy minimum. It shifts to smaller values of $q_{[\overline{1}10]}$ and also the energy gain relative to the FM state decreases. Analysis of the impact of hydrogenation on exchange parameters (see Note~2 in the SI) reveals a reduction in the intra-plane antiferromagnetic exchange parameter $J_{02}$, connecting spins along the [$\overline{1}10$], which becomes ferromagnetic. This is evident from the positive value of the spin stiffness at high charge/hydrogen concentrations, see Note~2 in the SI. In absence of additional interactions, the magnetic order would be ferromagnetic. However, a spin-spiral state can emerge solely due to a significant interface DMI competing with FM exchange whose strength is finely tuned by the appropriate H concentration. We find that the DMI is only weakly influenced by charging and thus is expected to retain the same chirality of the pristine system, see Note~3 in the SI. The weak effect of H on the DMI is consistent with prior observations~\cite{Chen2021}. As laboratory magnetic fields can compete with the DMI, such DMI-spirals could then transition to single skyrmions. 

We conducted H-adsorption experiments in order to assess the DFT prediction that the exchange-frustrated, short-period and stiff 2\,ML Fe/Ir(110) spin spiral can be transformed into a longer-period, DMI-dominated, and tunable spin spiral. H$_2$ was dissociatively adsorbed at 100\,K to submonolayer coverage. The sample was subsequently cooled to 1.7\,K in the STM. For details, see Note~4 and Note~5 in the SI. 

%Figure 2
\begin{figure*}[ht!]
\includegraphics[width=\textwidth]{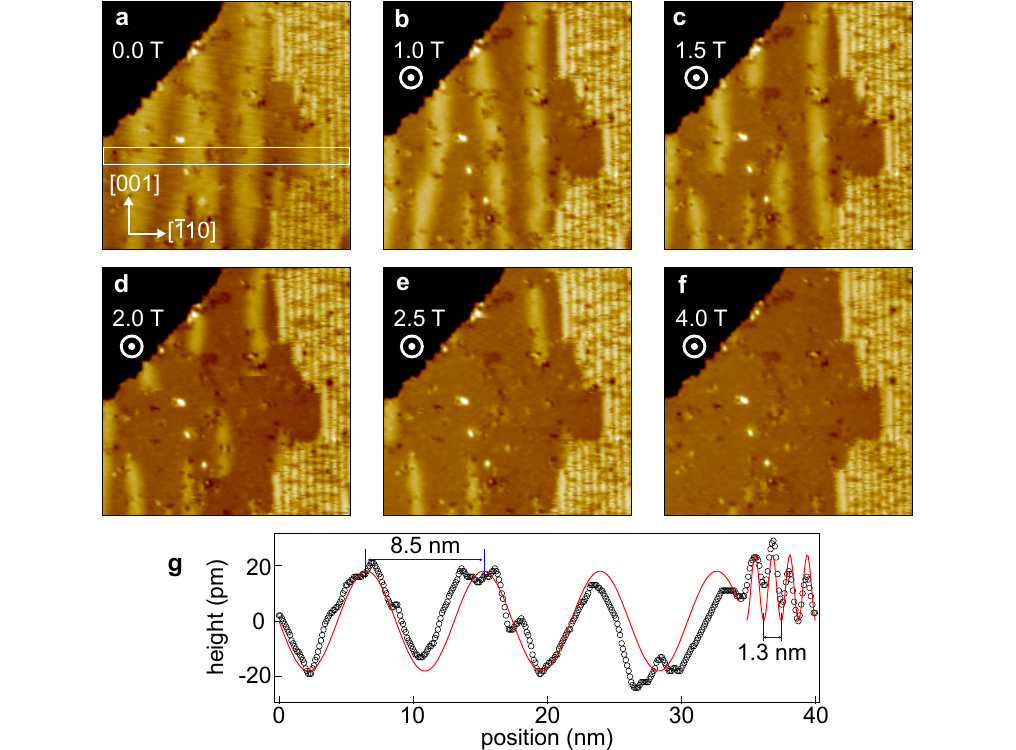}
\caption{Field induced unwinding of spin spirals in the dense H adsorption phase on 2\,ML Fe/Ir(110). Spin-polarized STM data obtained using a magnetically soft Fe-coated W tip with $\textbf{m}_\mathrm{tip}$ normal to surface. (a)-(f) Constant-current STM topographs after H$_2$ exposure at 100\,K with indicated increasing out-of-plane magnetic fields. Images with: $V_\mathrm{b} = 100$~mV, $I_\mathrm{set} = 1$~nA, and image size 40\,nm\,$\times$\,40\,nm. (g) Averaged line profile (black circles and line) taken within the white rectangle in (a); Red curve: sinusoidal curves with periods given in the graph.
}
\label{field_HFe}
\end{figure*}

We find that H adsorbs in two coexisting phases on 2\,ML Fe/Ir(110): (i) a dilute phase not changing the magnetic structure of the spin spiral described above and in Ref.~\onlinecite{Knispel2025} and identifiable only through increased noise induced by H diffusion and (ii) a dense H adsorption phase which is lower in apparent height by about 20\,pm and features a different electronic structure (see Note~5 in the SI). Additionally, the dense phase can be transformed to the dilute phase via voltage pulses or scanning with high voltage bias. This manipulation is displayed in Note~6 in the SI. The spin-polarized STM topograph of Figure~\ref{field_HFe}a taken with a soft magnetic tip displays a 2\,ML Fe/Ir(110) area with the two phases coexisting. On the right, the dilute H-phase can be recognized while on the left the dense H-phase can be recognized through the longer wavelength of the new spin texture. The profile in Figure~\ref{field_HFe}g obtained by averaging the scan lines in the white rectangle of Figure~\ref{field_HFe}a makes the different periodicities of the magnetic texture quantitative. In the dense H phase, the wavelength $\lambda_\mathrm{H} \approx 8.5$\,nm is nearly 7 times larger than the pristine system. The spin texture of the new ground state, determined similarly to Ref.~\onlinecite{Knispel2025}, is again a right-handed Néel spin spiral, see Note~7 of the SI. 

With increasing out-of-plane magnetic field $B$, the spin texture within the dense H-adsorption phase changes substantially as highlighted by the sequence of STM topographs of Figure~\ref{field_HFe}b-e. For $B=1.0$\,T the dark areas expand and the bright areas shrink. The spin spiral becomes distorted by orienting the moments towards the applied magnetic field. Note that here, due to a negative tip spin polarization, the magnetic contrast is reversed relative to the usual convention (dark areas indicate parallel and bright areas antiparallel orientation of $\textbf{m}_\mathrm{s}$ and $\textbf{m}_\mathrm{tip}$). When $B$ is increased to 1.5\,T, as in Figure~\ref{field_HFe}c, some spin spiral stripes break into pieces signifying a topological transition in the spin texture. Further increase to $B = 2.0$\,T and $B = 2.5$\,T as in Figure~\ref{field_HFe}d and e, respectively, decomposes the stripes into isolated magnetic skyrmions (magnetization antiparallel to applied field) within a sea of spins aligned to the external field. The skyrmions are elongated along the [001]-direction and do not condense into circular shape. Finally, for $B = 4$\,T as shown in Figure~\ref{field_HFe}f, the hydrogenated area is field-polarized, \textit{i.e.}, the sample is entirely magnetized parallel to the external field. Noteworthy, the magnetic stripe contrast in the dilute H-phase does not show any response up to the highest applied field.

In short, our experimental observations confirm a hydrogen-adsorption induced transformation of a stiff, short wavelength spin spiral resulting from exchange frustration to a tunable, long wavelength spin spiral resulting mainly from the DMI. The latter can be broken into elongated skyrmions and eventually resolved into a single domain of field-aligned spins with increasing $B$.

\begin{figure}[ht!]
\includegraphics[width=0.45\textwidth]{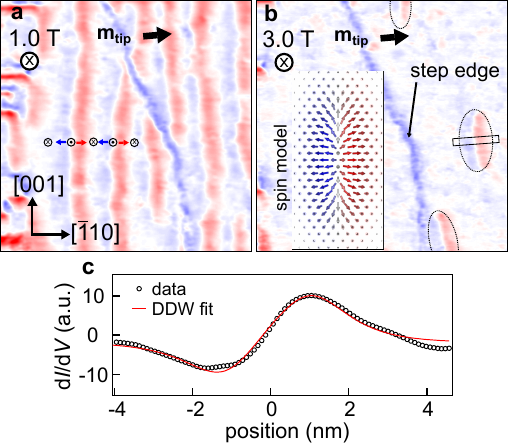}
\caption{Identification of magnetic skyrmions. (a), (b) $\mathrm{d}I/\mathrm{d}V$ maps of 2\,ML Fe/Ir(110) taken with a bulk Cr tip with fixed in-plane magnetization as indicated. The external magnetic field is normal to the surface into the sample. $B = 1.0$~T in (a) and $B = 3.0$~T in (b). The magnetic contrast is visualized with a red to blue color scale indicating parallel to anti-parallel alignment between $\textbf{m}_\mathrm{tip}$ and $\textbf{m}_\mathrm{s}$, respectively. The double lobes surrounding the magnetic stripes or skyrmions show their in-plane spin texture. Inset of (b) is the spin model simulation colored only along the tip magnetization experimental axis. $V_\mathrm{b} = 100$~mV, $I_\mathrm{set} = 1$~nA, and image size $50~\mathrm{nm} \times 50~\mathrm{nm}$. Spin model map: $6~\mathrm{nm} \times 12~\mathrm{nm}$
(c) Differential conductance linescan signal (black circles) averaged within the black rectangle in (b). The red line is a fit assuming the characteristic double domain wall of the skyrmion texture with Eq.\,\ref{ddw}.
}
\label{inplane}
\end{figure}

To uncover the nature of the skyrmion spin texture, we used an antiferromagnetic Cr tip with fixed in-plane magnetization, \textit{i.e.}, not affected by external magnetic field. Figure~\ref{inplane}a,b show $\mathrm{d}I/\mathrm{d}V$ maps of 2\,ML Fe islands separated by a step edge at different external fields of 1~T and 3~T, respectively. In Figure~\ref{inplane}a the low field state is presented. The in-plane sensitivity of the tip results in a different appearance of the spin spiral as compared to Figure~\ref{field_HFe}b. Here, only in-plane components present contrast, red is parallel and blue is antiparallel to the tip magnetization. The out-of-plane components are orthogonal to the tip magnetization and are imaged white. The broad white stripes display magnetization parallel and the narrow white stripes magnetization antiparallel to the external field, consistent with Figure~\ref{field_HFe}b.

%Figure 3

At the field of 3~T in Figure~\ref{inplane}b, we find isolated skyrmions on the surface. The skyrmions are now characterized by lobes of blue and of red color code, providing an additional piece of the spin texture. The spin orientation at the skyrmion center opposes the external field and gradually rotates into the plane with increasing distance, thereby establishing the skyrmionic nature of the structure. Further evidence for the skyrmionic nature is given by the quantitative analysis of the magnetic contrast. Following the work of Romming \textit{et al.}~\cite{Romming2015}, a cross section of a skyrmion through its center is approximated by a double domain wall width, \textit{i.e.}, by two overlapping domain walls through 
\begin{equation}
\phi_\mathrm{DDW}(x) = \sum_{+,-} \arcsin\left(\tanh\left(\frac{x \pm c}{w/2}\right)\right).
\label{ddw}
\end{equation}
Here, $\phi_\mathrm{DDW}(x)$ is the polar angle of \textbf{m}$_\mathrm{s}$ at position $x$, $2c$ defines the size of the skyrmion along the cross section and $w$ is the width of the domain walls. Taking into account the projection of \textbf{m}$_\mathrm{s}$ through the fixed \textbf{m}$_\mathrm{tip}$, the formula is fitted to the magnetic contrast taken along the shorter axis of the elliptical skyrmion within the black rectangle in Figure~\ref{inplane}b. Using $2c=2.12$~nm and $w=1.72$~nm, very good agreement between experimental data and the fit is obtained, as apparent in Figure~\ref{inplane}c. %\jfc{to SB, please check:} 
The quantitative agreement suggests that the double domain wall provides a reasonable description of the spin texture, which in the particular limits of spin stiffness ($A$), DMI vector ($D$), magnetic anisotropy ($K$), and field ($B$), is related to the skyrmionic texture.

Figure~\ref{inplane}b reveals that the elliptical skyrmion exhibits a size approximately twice as large along the [001] direction compared to the [$\overline{1}$10]-direction. To assess the origin of the skyrmion anisotropy in a quantitative manner, we map the dispersion curves of the of Figure~\ref{dft}a onto exchange parameters, $J_{ij}$, and nearest neighbors DMI, $D$ (for parameters, see Note~2 and Note~3 in the SI). Using the parameters for highest charge/H concentration, we simulated the spin order in the system via spin-dynamics simulations~\cite{Mueller2019,spiritCode}. As given in the inset of Figure~\ref{inplane}b, the spin model nicely reproduces the elliptical shape of the skyrmions, observed experimentally, with a similar aspect size of 2. The origin of the elliptical shape of the isolated skyrmions is the interplay between anisotropic exchange and DMI, giving rise to a stronger canting of spins along the [$\overline{1}$10]-direction and thus to the elliptic distortion.

In summary, DFT calculations predict that dense hydrogen adsorption  severely changes the balance between ferro- and antiferromagnetic Heisenberg interactions across different pairs of atoms, shifting the Yoshimori-spiral of 1.3~nm in a double layer of Fe on Ir(110) towards ferromagnetic  exchange interaction. This ferromagnetic interaction competes with the DMI to form a right-handed  DMI-stabilized spin spiral providing a route for a phase transition to a skyrmionic state under lab magnetic fields. Experimentally, controllable hydrogenation gives rise to coexistence of dilute and dense H adsorption phases on 2~ML Fe on Ir(110). While the spin texture in the dilute phase is unchanged, in the dense H phase the ground state changes to a spin spiral with wavelength of 8.5~nm. The spin spiral in the dense H phase is magnetically tunable and upon increasing external out-of-plane magnetic fields transforms into a skyrmionic state at 2.0~T. This state persists until 4.0~T, beyond which the magnetization becomes fully aligned with the field. 
Due to the low symmetry of the surface, the interplay of anisotropic exchange and DMI in the Fe/Ir(110) interface the skyrmions are ellipsoidal with longer and shorter axis oriented along the [001] and the [$\overline{1}$10] crystallographic directions, respectively. Moreover, through hydrogen desorption, the dense H phase can be transformed back to the phase with Yoshimori spiral. Thus, via adsorption and desorption of hydrogen the magnetic state can be toggled.

In the present case, the effective micromagnetic DMI is isotropic, while the elliptical shape of the skyrmions arises from anisotropic exchange interactions. The approach developed here can be extended to other systems with low C$_{2v}$ symmetry, where an anisotropic DMI may enable the formation of antiskyrmions on surfaces.
\\
\begin{acknowledgments}
	\noindent
	V.T.\ and S.B.\ thank Markus Hoffmann for his support in carrying out the DFT calculations. We acknowledge funding from Deutsche Forschungsgemeinschaft (DFG) through CRC 1238 (Grant No.\ 277146847, projects B06 and C01). S.B.\ and J.F.\ acknowledge financial support from DFG through projects BL~444/16-2 and FI~2624/1-1 within the SPP 2137 (Grant No.\ 462692705). S.B.\ acknowledges financial support from the European Research Council (ERC) under the European Union's Horizon2020 research and innovation program (Grant No.\ 856538, project “3D MAGiC”). G.B.\ gratefully acknowledges computing time granted through JARA-HPC on the supercomputer JURECA at Forschungszentrum Jülich.
\end{acknowledgments}

\subsection{Code availability.} The DFT code \texttt{FLEUR} used in this work is publicly accessible at \url{https://www.flapw.de}. The atomistic spin dynamics code \texttt{SPIRIT} used in this work is publicly accessible at \url{https://spirit-code.github.io}.

\bibliographystyle{apsrev4-2}
\bibliography{bib_HFeIr(110)}
\end{document}

\title{Sypplementary Information:\\
Hydrogen toggling between Yoshimori spin spirals and elliptical Dzyaloshinskii-Moriya skyrmions in Fe on Ir(110)}

\author{Timo Knispel}
\affiliation{II. Physikalisches Institut, Universit\"{a}t zu K\"{o}ln, Z\"{u}lpicher Str. 77, 50937 Cologne, Germany \looseness=-1}
\author{Vasily Tseplyaev}
\affiliation{Peter Gr\"{u}nberg Institut, Forschungszentrum J\"{u}lich and JARA, D-52425 J\"{u}lich, Germany \looseness=-1}
\affiliation{Physics Department, RWTH-Aachen University, D-52062 Aachen, Germany \looseness=-1}
\author{Gustav Bihlmayer}
\affiliation{Peter Gr\"{u}nberg Institut, Forschungszentrum J\"{u}lich and JARA, D-52425 J\"{u}lich, Germany \looseness=-1}
\author{Stefan Bl\"ugel}
\affiliation{Peter Gr\"{u}nberg Institut, Forschungszentrum J\"{u}lich and JARA, D-52425 J\"{u}lich, Germany \looseness=-1}
\affiliation{Physics Department, RWTH-Aachen University, D-52062 Aachen, Germany \looseness=-1}
\author{Thomas Michely}
\affiliation{II. Physikalisches Institut, Universit\"{a}t zu K\"{o}ln, Z\"{u}lpicher Str. 77, 50937 Cologne, Germany \looseness=-1}
\author{Jeison Fischer}
\email{jfischer@ph2.uni-koeln.de}
\affiliation{II. Physikalisches Institut, Universit\"{a}t zu K\"{o}ln, Z\"{u}lpicher Str. 77, 50937 Cologne, Germany \looseness=-1}

\maketitle
\vspace{-1cm}
\tableofcontents
\newpage
\subsection{Supplementary Note 1: First-principles calculations}

We performed vector-spin DFT calculations using the film version of the full-potential linearized augmented plane wave method~\cite{wimmer1981flapw} (FLAPW) as implemented in the FLEUR code~\cite{fleur2023} (www.flapw.de). By using the FLEUR code, we get access to the total energy, magnetic moment, and the atomic structure of non-collinear magnetic structures and spin spirals both with and without spin–orbit coupling (SOC).

The structural optimizations were performed for an 11-layer Ir(110) film terminated with two Fe layers on one of the surfaces, allowing the top four layers to relax. For the in-plane lattice parameters and the lower-lying Ir interlayer distances we used the optimized bulk lattice constants of 0.382~nm or 0.384~nm for the local density approximation (LDA)~\cite{Vosko:80} or mixed generalized gradient approximation (GGA)~\cite{DeSantis:07}, respectively. The muffin-ion radii of H, Fe, and Ir were $0.91$, $2.09$, and $2.33\,a_0$ and the plane-wave cutoff was $4.2\,a_0^{-1}$. For the relaxations, a regular $12 \times 8$ k-point grid was chosen. The relaxations are shown in Table~\ref{tab:relax}.

For the hydrogen-covered surfaces, several adsorption sites in the p$(1\times1)$ unit cell were tested. The most stable configuration was identified with H located in a bridge position between the topmost Fe atoms, as shown in the inset of Figure~1b in the main text. As the energy difference to all other tested positions was larger than $1$~eV, the same position was also chosen for smaller coverages in the p$(2\times2)$ unit cells. The relaxations are again listed in Table~\ref{tab:relax}.
\begin{table}[h]
    \centering
    \begin{tabular}{|c|rrrrr|}
          \hline 
                & 0.0 H & 0.0 H & 0.25 H & 0.75 H & 1.0 H \\
                & [LDA] & [GGA] &  [LDA] &  [LDA] & [LDA] \\
         \hline
         \hline
        H  - Fe &       &       & $-24$  & $-26$  & $-27$ \\
        Fe - Fe & $-19$ & $-16$ & $-20$  & $-18$  & $-11$ \\
        Fe - Ir & $ -7$ & $ -4$ & $ -8$  & $ -8$  & $-10$ \\
        Ir - Ir & $ -1$ & $ +1$ &        &        &       \\
        \hline
    \end{tabular}
    \caption{Relaxation of interlayer distances in percent of the theoretical bulk interlayer distance of Ir(110). Only the topmost four layers were allowed to relax. 0.0 H corresponds to the clean surface, 1.0 H to the fully hydrogen covered one. The intermediate occupations were calculated in a p$(2\times2)$ unit cell.}
    \label{tab:relax}
\end{table}

For the spin-spiral calculations, 3840~$\mathbf k_{\|}$-points in the full two-dimensional (2D) Brillouin-zone were used. Conical spin-spirals with an opening angle of 30 degree were calculated using the generalized Bloch theorem~\cite{Kurz:04} and the magnetic force theorem~\cite{Lezaic2013}. For calculating the Dzyaloshinskii-Moriya interaction (DMI), SOC was included in first-order perturbation theory~\cite{Heide:09}. For the calculation of the magnetic anisotropy, SOC was included self-consistently~\cite{Li:90}.

It is important to note that the dispersion of the spin-spiral is highly sensitive to atomic relaxations.  LDA gives smaller Fe-Ir interlayer distances than the GGA functional and leads to shorter spin-spirals with larger stabilization energy. This is mainly an effect of the structure, not of the choice of the exchange-correlation functional used to calculate the spin-spiral dispersion itself, although the choice of the exchange correlation type determines the details of the atomic structure. Nevertheless, the sense of rotation is not changed, and the wavelength remains in the same range.

\clearpage

\subsection{Supplementary Note 2: Effect of charge on the exchange parameters}

\begin{figure}[h]
  \begin{minipage}[b]{.5\linewidth}
    \centering
    \includegraphics[width=0.7\textwidth]{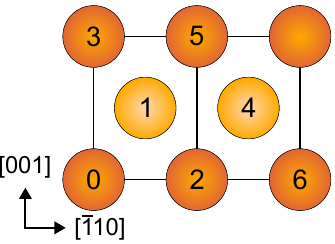}
    \captionof{figure}{Ball model double layer Fe.}% \caption{Figure caption}
    \label{fig_ball_model}
  \end{minipage}\hfill
  \begin{minipage}[b]{.5\linewidth}
    \centering
    \begin{tabular}{|c|cccc|}
    \hline
    & Clean & +0.1e & +0.2e & +0.3e \\
    \hline
    J$_{01}$ & 51.8 & 51.8 & 51.8 & 51.8 \\
    J$_{02}$ & -20.5 & -5.6 & 7.7 & 19.7 \\
    J$_{03}$ & -27.1 & -23.3 & -17.1 & -11.8 \\
    J$_{06}$ & -7.4 & -9.3 & -10.2 & -10.2 \\
    \hline
    \end{tabular}
    \captionof{table}{Exchange constants (meV/Fe atom).}
    \label{tab_exchange_constants}
  \end{minipage}
\end{figure}

Exchange constants $J_{0i}$ between the 0-th and the i-th neighbor (see  neighboring in Figure~\ref{fig_ball_model}) obtained from fitting the energy dispersion of homogeneous spin spirals calculated by means of scalar-relativistic DFT for Fe/Ir(110) as shown in Figure~1a (main text).

\begin{table}[h]
    \centering
    \begin{tabular}{|c|ccc|cccc|cccc|}
        \hline
        & \multicolumn{3}{ c|}{} & clean & +0.1e & +0.2e & +0.3e & clean & +0.1e & +0.2e & +0.3e \\ 
        \hline
        & m & n$_{[\bar{1}10]}$ & n$_{[001]}$ & \multicolumn{4}{c|}{m*J$_{0i}$(n$_{[\bar{1}10]})^2$} & \multicolumn{4}{c|}{m*J$_{0i}$(n$_{[001]})^2$} \\        
        \hline
        J$_{01}$ & 2 & 1 & 1 & 103.6 & 103.6 & 103.6 & 103.6 & 103.6 & 103.6 & 103.6 & 103.6\\
        J$_{02}$ & 1 & 2 & 0 & -82.0 & -22.4 & 30.8 & 78.8 & - & - & - & -  \\
        J$_{03}$ & 1 & 0 & 2 & - & - & - & - & -108.4 & -93.2 & -68.4 & 47.2 \\
        J$_{06}$ & 1 & 4 & 0 & -118.4 & -148.8 & -163.2 & -163.2 & - & - & - & - \\
        \hline
        \multicolumn{4}{|c|}{spin stiffness at $q=0$} & \cellcolor{red!25}$-$96.8 & \cellcolor{red!25}$-$67.6 & \cellcolor{red!25}$-$28.8 & \cellcolor{blue!25}19.2 & \cellcolor{red!25}$-$4.8 & \cellcolor{blue!25}10.4 & \cellcolor{blue!25}35.2 & \cellcolor{blue!25}56.4 \\
        
         \hline
    \end{tabular}
    \caption{Contributions $m J_{0i} (n_{\hat{\mathbf{q}}})^2$ to the curvature of the energy dispersion of the spin spiral at $q=0$, $\alpha_{\hat{\mathbf{q}}} = \sum_i m J_{0i} (n_{\hat{\mathbf{q}}})^2$, with $\hat{\mathbf{q}} = [\overline{1}10]$\, or $[001]$, calculated taken into account the number of similar neighbors given by the multiplicity (m), exchange constants $J_{0i}$ from Table\,\ref{tab_exchange_constants}, and an integer factor $n_{\hat{\mathbf{q}}}$ multiplying the nearest neighbor distance ($a_{nn}$) along a specific direction. The sign of $\alpha$ indicates whether the FM state is stable ($\alpha > 0$, cell colored blue) or unstable ($\alpha < 0$, cell colored red). To convert to micromagnetic spin stiffness parameters, with energy expressed in meV/nm$^2$ and $q$ in 1/nm, the values have to be multiplied by $\sqrt{2}/8$ ($\sqrt{2}/4$) for the $[\overline{1}10]$ ($[001]$) direction.}
    \label{tab_spin_stiffness}
\end{table}

\begin{table}[h]
    \centering
    \begin{tabular}{|cccc|c|}
        \hline
        \multicolumn{5}{|c|}{Magnetic Moments ($\mu_B$)} \\
        \hline
        clean & +0.1e & +0.2e & +0.3e & +1H\\
        \hline
        2.59 & 2.52 & 2.44 & 2.36 & 2.37 \\
        \hline
    \end{tabular}
    \caption{Averaged magnetic moments of the two Fe layers under different conditions.}
    \label{tab:magnetic_moments}
\end{table}

\clearpage

\subsection{Supplementary Note 3: Effect of charge on the DMI}

\begin{figure}[h]
\centering
\includegraphics{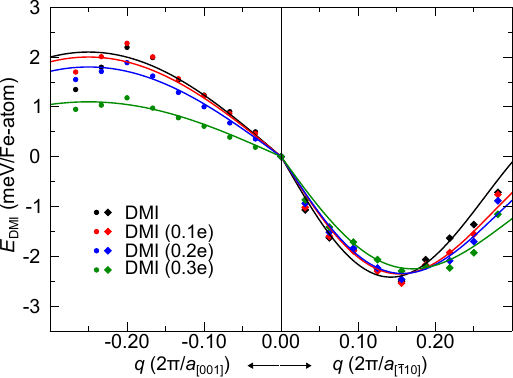}
\caption{DFT calculated energetics of the DMI.
Energy dispersion of clean 2\,ML Fe/Ir(110) and with virtually additional charge of 0.1e, 0.2e and 0.3e along two orthogonal directions. DMI is mostly unchanged with the additional charge.
}
\label{fig_chargedFe}
\end{figure}

\clearpage

\subsection{Supplementary Note 4: Experimental methods}

The experiments were performed in an ultra-high vacuum (UHV) system (base pressure $<$ 2$\times$10$^{-10}$~mbar) equipped with a sample preparation chamber and an STM operating at 1.7~K with superconducting magnets producing in the STM sample position a magnetic field of up to 9~T normal to the sample surface and 2~T in any direction.

Preparation details can be found in Ref.~\cite{Knispel2025}. Ir(110) is sputtered with 1~keV Ar$^+$ and annealed to 1500~K. During cooling from 1200~K, the surface is exposed to an oxygen partial pressure of $1 \times 10^{-7}$\,mbar to prevent the formation of the nanofacet reconstruction. Then, the chemisorbed oxygen is titrated through exposure to $1 \times 10^{-7}$\,mbar hydrogen at 470\,K until clean Ir(110)-$(1 \times 1)$ results. 

Fe amounts ranging from 0.4\,ML to 1.4\,ML Fe were deposited by e-beam evaporation onto clean Ir(110)-$(1 \times 1)$ at 470~K. Here, ML denotes monolayer and 1\,ML corresponds to the surface atomic density of Ir(110)-$(1 \times 1)$, which is $9.6 \times 10^{18}$\,atoms/m$^2$.

Hydrogenation of Ir(110) with Fe islands was carried out at about 100\,K with a 50\,s exposure to an H$_2$ partial pressure of $5.2 \times 10^{-9}$~mbar. The pressure is corrected for the low ion gauge sensitivity of H$_2$. This exposure equates to 0.45 Langmuir. Assuming a sticking coefficient around 1 for Ir(110), as deduced by Ibbotson \textit{et al}.~\cite{Ibbotson1980}, we estimate a coverage of nearly 1 hydrogen atom per surface site.

For spin-polarized STM measurements, we used Fe-coated W and bulk Cr tips. To obtain an Fe-coated W tip, an electrochemically etched W tungsten tip was heated in UHV up to 2400\,K for 2\,s and subsequently coated by 20 layers of Fe. After deposition the Fe-coated W tip was annealed at 1000\,K for 5\,s. Fe-coated tips display soft magnetization, i.e., their magnetization follows the direction of the external field~\cite{Phark2013}. A Cr bulk tip was obtained by electrochemical etching of a rectangular piece of Cr, which was mildly heated in UHV up to 1000\,K for 2\,s. The magnetization of Cr tips is hard and unaffected by the external field.
We detect the tunnel current $I$($V$) and the differential conductance ($\mathrm{d}I/\mathrm{d}V$) simultaneously using a lock-in technique with a modulation bias voltage of $20$~mV at a frequency $6.662$~kHz.

\clearpage
\subsection{Supplementary Note 5: Hydrogenation of 2\,ML Fe islands on Ir(110)}

\begin{figure}[h]
\centering
\includegraphics[width=\textwidth]{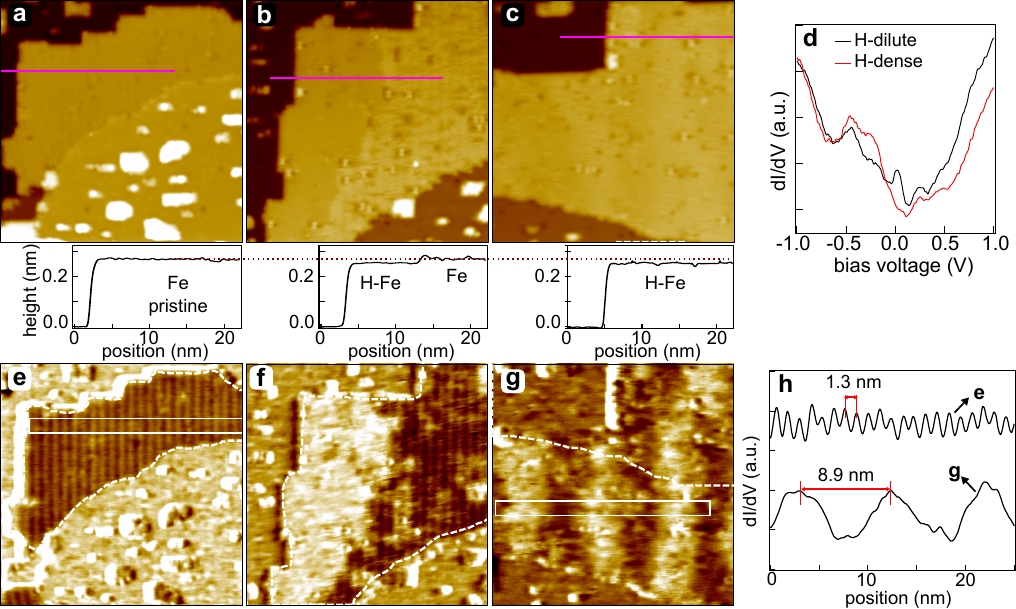}
\caption{Hydrogenation induced change of spin texture.
Constant-current STM images (a) of pristine Fe islands on Ir(110), (b) after exposure to 0.22~Langmuir, and (c) after exposure to 0.45~Langmuir of H$_2$. Line profiles of individual islands along magenta lines are shown below the respective topographs. The comparison of heights reveals that upon hydrogenation a new phase with lower apparent height is formed, the difference is measured to be $\approx$ 0.01~nm.
(d) $\mathrm{d}I/\mathrm{d}V$ spectra taken on the dilute H-phase Fe (black) and on the dense H-phase Fe (red) shows the different electronic structures. The main change is close to the Fermi energy, as the characteristic peak is suppressed in the dense H-phase.
(e),(f) and (g) Spin-polarized $\mathrm{d}I/\mathrm{d}V$ maps of the same areas as (a),(b), and (c), respectively.
The modulation pattern visible in (e) is restricted to a part of the Fe island in (f). In (g) the short period modulation pattern is suppressed and a new modulation with large period is visible. 
(h) Averaged $\mathrm{d}I/\mathrm{d}V$ line profiles taken within the rectangles in (e) and (g). 
For all images $V_\mathrm{b} = 100$\,mV, $I_\mathrm{set} = 1$\,nA, and 28\,nm\,$\times$\,28\,nm.
 }
\label{figS5}
\end{figure}

\clearpage

\subsection{Supplementary Note 6: Switching between H-dense and H-dilute surface phases}

\begin{figure}[h]
\centering
\includegraphics[]{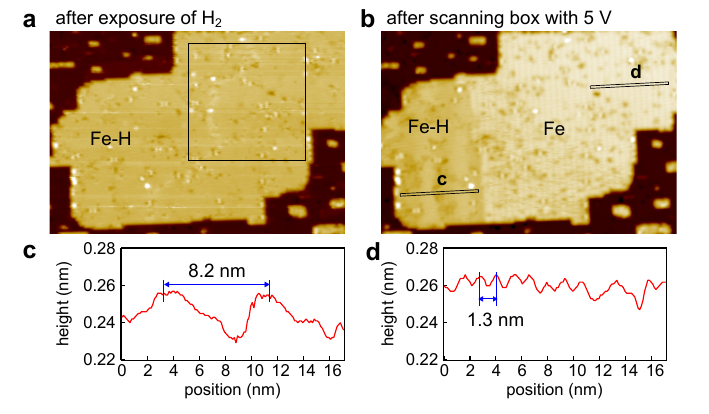}
\caption{Recovery of H-dilute from H-dense phase by STM scanning.
(a) STM image of a 2\,ML Fe island on Ir(110) taken after exposure to 0.2~L of H$_2$. Surface is covered by the H-dense phase. (b) STM image of the same Fe island taken after the indicated box in (a) was scanned with bias of 5~V. (c,d) Line profiles of H-dense and H-dilute phases, respectively, taken along the marked lines in (b). For both images $V_\mathrm{b} = 100$\,mV, $I_\mathrm{set} = 1$\,nA, and 60\,nm\,$\times$\,40\,nm.
}
\label{toggling}
\end{figure}

\clearpage

\subsection{Supplementary Note 7: Chirality of the spin spiral in hydrogenated Fe islands.}

\begin{figure}[h!]
\centering
\includegraphics[width=0.85\textwidth]{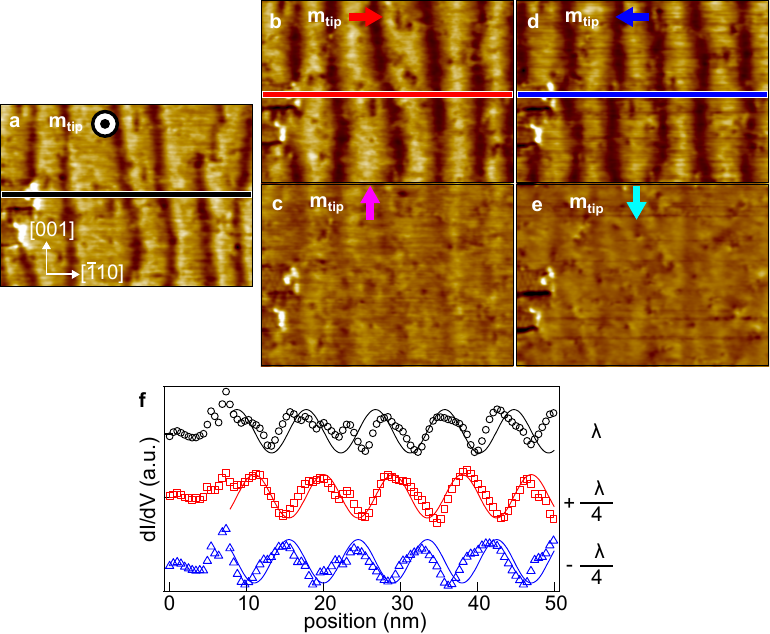}
\caption{Identification of chirality of the spin spiral in H-Fe.
Spin-polarized $\mathrm{d}I/\mathrm{d}V$ maps and profiles recorded on a 2\,ML Fe island. The tip magnetization orientation $m_\mathrm{tip}$ is given in each sub figure. Magnetic stripe modulation is observed within the 2\,ML Fe region for (a) upwards out-of-plane tip magnetization and (b),(d) in-plane tip magnetization along [$\overline{1}10$] and [1$\overline{1}0$]-direction.
(c),(e) The stripe contrast vanishes for tip magnetization along the [001]- and the [00$\overline{1}$]-directions, respectively.
(h) Averaged $\mathrm{d}I/\mathrm{d}V$ line profiles taken within the white rectangles in (a) (black circles), (b) (red squares), and (d) (blue triangles). The color code of the profiles matches with the color code used in the $\mathrm{d}I/\mathrm{d}V$ maps for tip magnetization. Black: out-of-plane, red along the in-plane [$\overline{1}$10]-direction, and magenta along the in-plane [001]-direction. The profiles are taken at the same location, as indicated by the unchanged position of the defect on the left side of the topographs. For \textbf{m}$_\mathrm{tip}$ along [$\overline{1}10$], red (or [1$\overline{1}0$], blue) the wave pattern is shifted forward (backward) by $+(-)\lambda/4$ with respect to \textbf{m}$_\mathrm{tip}$ normal to the surface (black).
For maps (a-e) $V_\mathrm{b} = 100$\,mV, $I_\mathrm{set} = 1$\,nA, and image size is 50\,nm\,$\times$\,36\,nm.
}
\label{hydrogenation}
\end{figure}

\clearpage

\subsection{Supplementary Note 8: Field dependence of the spin texture of hydrogenated Fe film measured with a hard Cr tip}

\begin{figure}[h]
\centering
\includegraphics[]{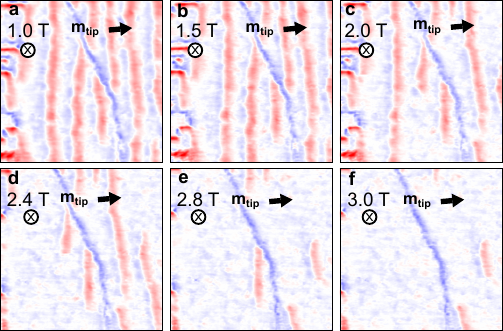}
\caption{$\mathrm{d}I/\mathrm{d}V$ maps of hydrogenated 2\,ML Fe/Ir(110) taken with a bulk Cr tip with fixed in-plane magnetization as indicated. The external magnetic field is normal to the surface into the sample. $B = 1.0$\,T in (a), $B = 1.5$\,T in (b), $B = 2.0$\,T in (c), $B = 2.4$\,T in (d), $B = 2.8$\,T in (e), and $B = 3.0$\,T in (b). The magnetic contrast is visualized with a red to blue color scale indicating parallel to anti-parallel alignment between $\textbf{m}_\mathrm{tip}$ and $\textbf{m}_\mathrm{s}$, respectively. The double lobes surrounding the magnetic stripes or bubbles show their in-plane spin texture. The overall image is blueish due to a small $m_\mathrm{tip}$ component out-of-plane. $V_\mathrm{b} = 100$\,mV, $I_\mathrm{set} = 1$\,nA, and image size 50\,nm\,$\times$\,50\,nm.
}
\label{inplane}
\end{figure}
\clearpage

\bibliography{bib_HFeIr(110)}